\documentclass{appolb}
\pdfoutput=1

\usepackage{epsfig}
\usepackage{wrapfig}
\usepackage{amsmath}
\usepackage{amssymb}
\usepackage{amsthm}
\usepackage{dcolumn}
\usepackage{graphics}
\usepackage{graphicx}
\usepackage{slashed,epsfig}
\usepackage{longtable}
\usepackage{color}

\definecolor{darkgreen}{rgb}{0,0.5,0}
\definecolor{purple}{rgb}{0.5,0,0.5}
\definecolor{nblue}{rgb}{0.0,0.0,0.50}
\definecolor{scarlet}{rgb}{1.0,0.2,0}
\usepackage[colorlinks=true, pdfstartview=FitV, linkcolor=purple, citecolor= purple, urlcolor=blue]{hyperref}

\newcommand{\nn}{\nonumber}
\newcommand{\beq} {\begin{equation}}
\newcommand{\eeq} {\end{equation}}
\newcommand{\beqa} {\begin{eqnarray}}
\newcommand{\eeqa} {\end{eqnarray}}

\newcommand{\wrt}{{\it wrt.\ }}

\newcommand{\as}{\alpha_s}

\newcommand{\chat}{\hat{\bs{\ell}}}

\newcommand{\ieps}{i\varepsilon}

\newcommand{\order}[1]{${\cal O}\left(#1 \right)$}
\newcommand{\morder}[1]{{\cal O}\left(#1 \right)}
\newcommand{\eq}[1]{(\ref{#1})}

\newcommand{\dpve}{\frac{d^3\pv}{(2\pi)^3\,2E_{p}}}
\newcommand{\rket}{\ket{0}_{R}}
\newcommand{\rbra}{{_{R}\bra{0}}}

\newcommand{\inv}[1]{\frac{1}{#1}}
\newcommand{\ket}[1]{\vert{#1}\rangle}
\newcommand{\bra}[1]{\langle{#1}\vert}

\newcommand{\com}[2]{\left[{#1},{#2}\right]}
\newcommand{\acom}[2]{\left\{{#1},{#2}\right\}}

\newcommand{\la}{\Lambda}
\newcommand{\bs}[1]{\boldsymbol{#1}}

\newcommand{\tpsi}{{\tilde{\psi}}}

\newcommand{\mA}{\mathcal{A}}

\newcommand{\mD}{\mathcal{D}}

\newcommand{\xv}{{\bs{x}}}
\newcommand{\yv}{{\bs{y}}}
\newcommand{\pv}{{\bs{p}}}
\newcommand{\kv}{{\bs{k}}}

\newcommand{\gv}{\bs{\gamma}}

\newcommand{\nv}{\bs{\nabla}}

\newcommand{\qu}{{\rm q}}
\newcommand{\qb}{{\rm\bar q}}

\newcommand{\halft}{{\textstyle \frac{1}{2}}}
\newcommand{\quart}{{\textstyle \frac{1}{4}}}

\begin{document}

\title{Hadron Structure %
\thanks{Presented at the 50th Cracow School of Theoretical Physics, 9-19 June 2010 in Zakopane, Poland.}}
\author{Paul Hoyer
\address{Department of Physics and Helsinki Institute of Physics\\ POB 64, FIN-00014 University of Helsinki, Finland}}
\maketitle
\begin{abstract}
I discuss a Born ($\hbar\to 0$) approximation of hadrons, motivated by a general feature of the data: The spectra of hadrons reflect their valence ($q\bar q$ or $qqq$) constituents, whereas hard scattering reveals a prominent sea quark distribution. Why do the sea quark d.o.f's not imply a richer spectrum? I look for an approach that can reconcile the quark and parton model descriptions of hadrons, and consider how this physics could emerge from the QCD Lagrangian. The possibilities are reduced by insisting that the approximation should be simple, yet adhere to the rules of quantum field theory. One might suspect that no such method exists -- but the Born approximation presents itself. The description of relativistic bound states that it brings has interesting features which merit further exploration.
\end{abstract}

\vspace{-12cm}
{\par\raggedleft \texttt{HIP-2010-25/TH}\par}
\vspace{12cm}
  
\section{Introduction}

Hadrons have an intriguing double nature, which seems paradoxical based on our experience with non-relativistic bound states. On the one hand, meson and baryon quantum numbers are successfully classified in terms of their valence quark ($q\bar q,\ qqq$) constituents \cite{pdg}. Their gluon degrees of freedom are not evident, as shown by inconclusive searches for glueballs and hybrids. The quark model is generally successful in explaining masses and magnetic moments (with the notable exception of the pion mass). 

%
\begin{figure}\label{fig1}
\includegraphics[width=12.5cm]{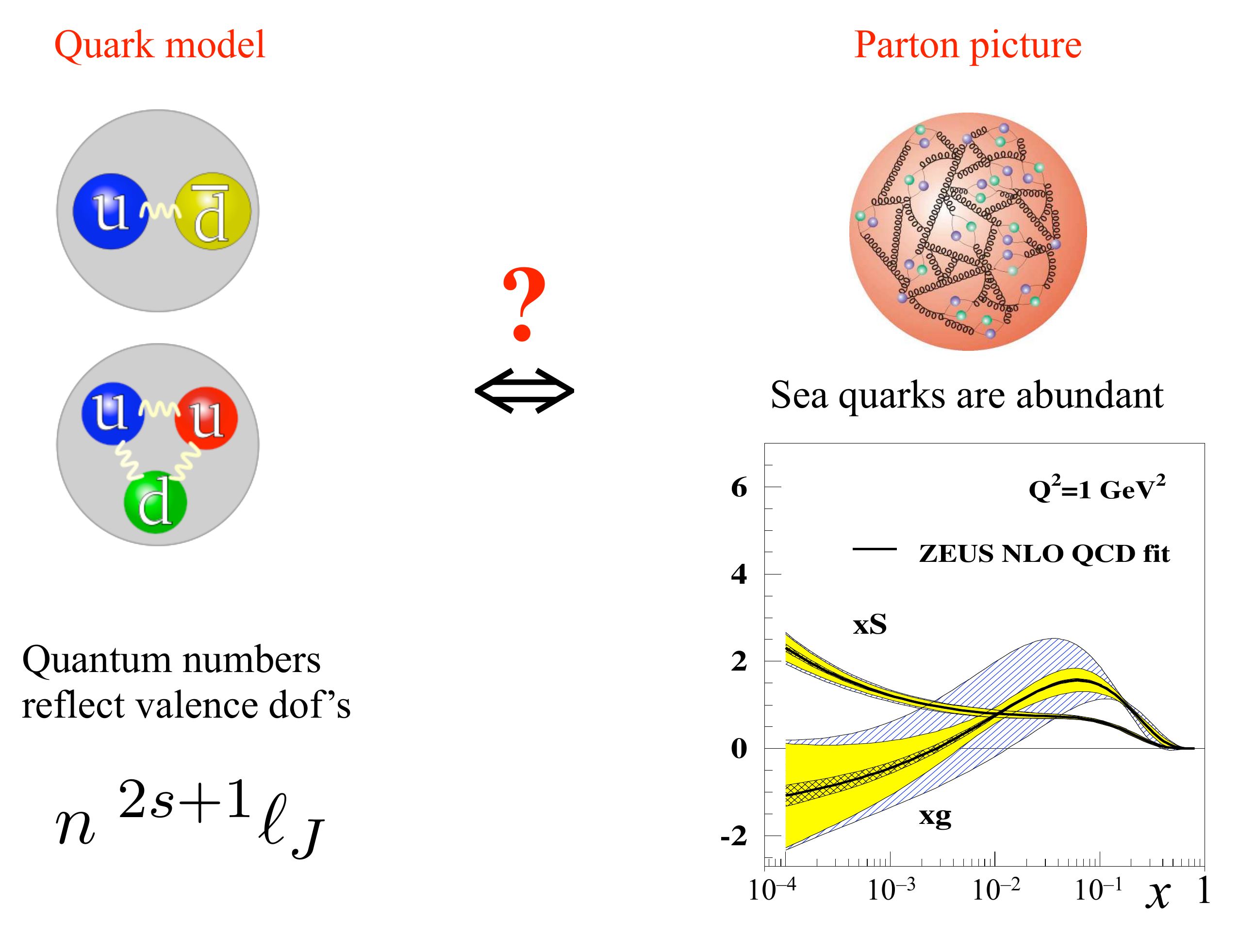}
\caption{\label{Fig1}The Quark model (left) describes the hadron spectrum solely in terms of the valence constituents. The parton model (right) views hadrons as having important gluon and sea quark components. The sea quark and gluon distributions at $Q^2=1$~GeV$^2$ (lower right) are taken from \cite{cooper}.}
\end{figure}
%

On the other hand, deep inelastic scattering shows that there are many gluons and sea quarks in hadrons. As seen in Fig.~1 there is an abundance of sea quarks even at the lowest $Q^2$ to which parton distributions may be reasonably extrapolated \cite{cooper}. Since the proton is an ultra-relativistic bound state it is hardly surprising that it contains many light quark pairs.

So why do the sea quark degrees of freedom not contribute to the spectrum? This is most likely related to their relativistic nature. Time-ordered perturbation theory shows that the intermediate states which contribute to scattering processes are frame-dependent. Fig.~2 illustrates how a covariant Feynman diagram describing Coulomb scattering from an external source breaks into two time-ordered contributions. The relative weight of the second ``$Z$''-diagram, which describes a particle pair fluctuation, depends on the frame. In the limiting case of the infinite momentum frame the $Z$-diagram is absent altogether. The Fock state structure of bound state wave functions must be similarly frame dependent. However, the quantum numbers of the bound state cannot change under boosts. Particle pair fluctuations apparently do not ``count'' as constituent degrees of freedom.  

%
\begin{figure}[t]\label{fig2}
\includegraphics[width=12.5cm]{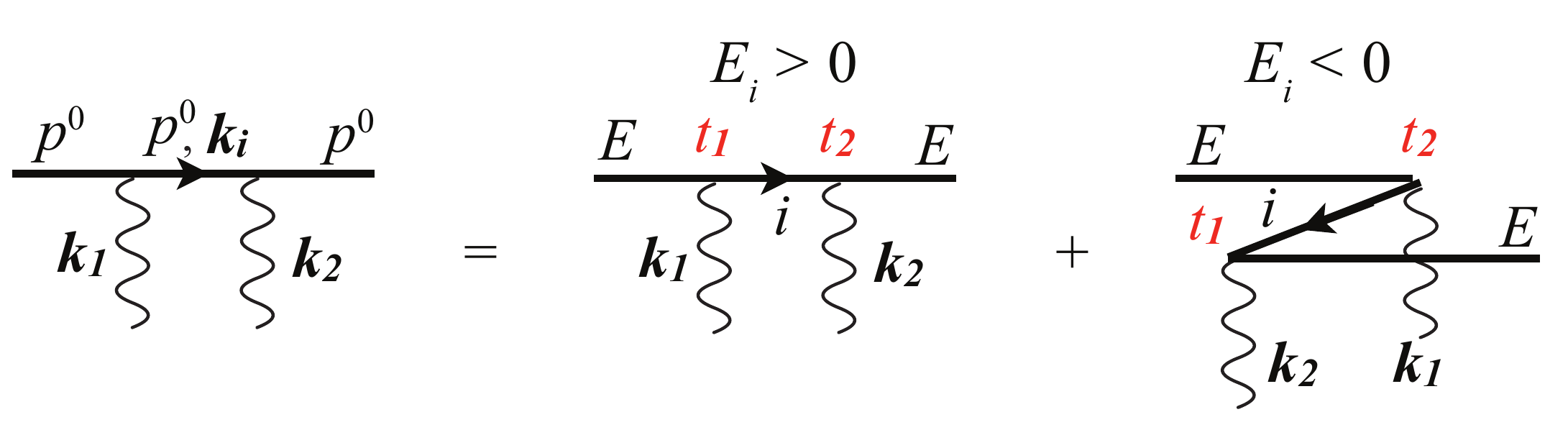}
\caption{Coulomb scattering from a time-independent source preserves the energy component $p^0=E$ of the scattering particle (covariant Feynman diagram on the left). In $(t,\pv)$ space the amplitude is given by the sum of two time-ordered diagrams (right). The second ``$Z$''-diagram has a three-particle intermediate state.}
\end{figure}
%

%
\begin{figure}[h]\label{fig3}
\includegraphics[width=12.5cm]{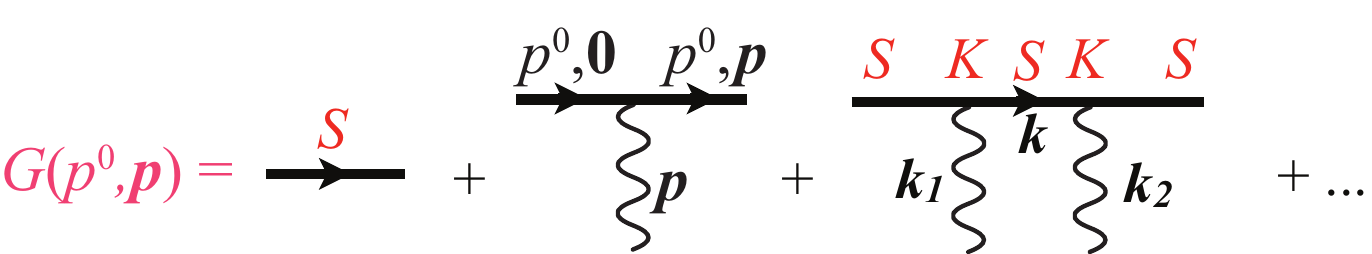}
\caption{The Green function $G$ formed by multiple interactions ($K$) of an electron propagating ($S$) in a static Coulomb potential. The initial and final electron momenta are denoted $(p^0,\bs{0})$ and $p=(p^0,\pv)$.}
\end{figure}
%

The above argument is rather vague. It may, however, be formulated more precisely for electrons bound in a time-independent Coulomb potential \cite{Hoyer:2009ep,Brodsky:2010zk}. The Green function $G(p^0,\pv)$ of the electron shown in Fig.~3 has poles at the bound state energies, say at $p^0= E_R$,
\beq\label{dseq}
G(p^0,\pv) = S+SKS+SKSKS+\ldots = S+SKG = \frac{R(E_R,\pv)}{p^0-E_R}+\ldots
\eeq
The last equality displays this pole contribution, whose residue $R(E_R,\pv)$ satisfies the Dirac equation in momentum space.  

Due to the static nature of the potential the interactions transfer only 3-momentum, conserving the energy component $p^0$ of the electron momentum. Hence for $p^0 >0$ none of the intermediate electron propagators $S(p^0,\kv)$ depend on the $\ieps$ prescription at $p^0 = -E_k$, with $E_k=\sqrt{\kv^2+m^2}$. The Green function $G(p^0,\pv)$ in \eq{dseq} is thus exactly the same whether a Feynman or Retarded prescription is used for the intermediate electron propagators,
\beq\label{frprop}
S_{F/R}(p^0,\kv) \equiv i\frac{p^0\gamma^0-\kv\cdot\gv+m}{(p^0-E_{k}+\ieps)(p^0+E_{k}\mp\ieps)}
\eeq
In particular, the energy $E_R$ of the bound state is unaffected by the $\ieps$ prescription.

Expressing each intermediate propagator $S(p^0,\kv)$ in \eq{dseq} as a Fourier transform of the time-ordered $S(t,\kv)$ reveals the intermediate states that contribute to the electron's bound state wave function at an instant of time $t$. Since Feynman propagators have support for $t<0$ they give rise to $Z$-diagrams as in Fig.~2. Retarded propagators on the other hand only allow forward motion in time, 
\beqa\label{srt}
S_{R}(t,\kv) = \frac{\theta(t)}{2E_{k}}\left[(E_{k}\gamma^0-\kv\cdot\gv+m)e^{-iE_{k}t} +(E_{k}\gamma^0+\kv\cdot\gv-m)e^{iE_{k}t}\right]
\eeqa
and give no $Z$-diagrams. Hence there is no pair production and the bound state appears to contain a single electron, which can have either positive or negative energy. The distribution of this electron is given by the standard bound state wave functions determined by the Dirac equation. 

Even though this was a rather trivial example, the conclusion is interesting. The same bound state can be described using two quite different equal-time wave functions. The ``true'' picture is given by the Feynman propagator and leads to a wave function with Fock states containing any number of pairs. The momentum distributions of all those pairs should be specified to fully describe the bound state. On the other hand, the same bound state is described by the usual Dirac single particle wave function when retarded propagators are used.

\section{The $\hbar$ expansion}

The possibility to describe relativistic Dirac bound states using a single particle wave function, in spite of the multiple pairs they contain, is analogous the dual nature of hadrons depicted in Fig.~1. Specifically, the Dirac spectrum reflects only the degrees of freedom of the single electron. Can such a description be extended to bound states formed by the interactions of relativistic particles, without an external potential?

The use of retarded propagators in the Dyson-Schwinger equation \eq{dseq} was allowed due the absence of loops. All of the diagrams in Fig.~3 are tree-level, Born diagrams. Does the concept of ``Born approximation'' exist for bound states? For scattering amplitudes the tree approximation is generally a good first approximation. Similarly, adding loop corrections to the propagators and vertices in Fig.~3 should not substantially change the nature of the Dirac bound states.

Tree diagrams are generally regarded as the leading contribution to scattering amplitudes in the limit where the Planck constant $\hbar \to 0$ \cite{hbarloops}. This is also the limit where one expects the laws of classical physics to apply. The fact that there are no classical bound states would suggest that there can be no Born level bound states, either. However, the relation between the $\hbar \to 0$ limit and classical physics is not straightforward \cite{Holstein:2004dn}, as I shall next demonstrate for the harmonic oscillator \cite{Brodsky:2010zk}.

The non-relativistic propagation of a particle from $(t_{i},x_{i})$ to $(t_{f},x_{f})$ in a potential $V(x)=\halft m\omega^2 x^2$ is given by the path integral
\beqa\label{ho}
\mA(x_{i},x_{f};t_{f}-t_{i}) &=& \int[\mD x(t)]\exp\left[\frac{i m}{2\hbar}\int_{t_{i}}^{t_{f}}dt ({\dot x}^2-\omega^2 x^2)\right]\nn\\ &=& \int[\mD\xi(t)]\exp\left[\frac{im}{2}\int_{t_{i}}^{t_{f}}dt ({\dot \xi}^2-\omega^2 \xi^2)\right]
\eeqa
In the second equality the explicit dependence on $\hbar$ was removed by scaling the coordinates as $\xi\equiv x/\sqrt{\hbar}$ (the Jacobian is irrelevant for this discussion.). The full quantum mechanical bound state structure of the harmonic oscillator thus persists as $\hbar \to 0$ in terms of the scaled variables $\xi$. The usual argument that the rapid oscillation of $\exp(i{\cal S}/\hbar)$ selects classical paths for which the action ${\cal S}$ is stationary fails for propagation over distances of \order{\sqrt{\hbar}}, where the variation of the action itself is of \order{\hbar}. It is only propagation between fixed ($\hbar$-independent) positions $x_{i},x_{f}$ which becomes classical in the $\hbar \to 0$ limit, since that involves highly excited levels ($E_{n}=\hbar\omega(n+\halft)$ with $n$ of order $1/\hbar$).

For the harmonic oscillator the concept of a Born approximation for bound states is rather trivial, since the dynamics in the rescaled variables is independent of $\hbar$. In gauge field theory the $\hbar$ dependence is more interesting. Rescaling the fields of the QED action gives
\beqa\label{qedact0}
{\cal S}_{QED}/\hbar &=& 
\inv{\hbar}\int d^4x\big[\bar\psi(i\slashed{\partial}-\tilde e\slashed{A}-\tilde m)\psi - \quart F_{\mu\nu}F^{\mu\nu}\big]\label{qedact}\\
&=& \int d^4x\Big[\bar\tpsi(i\slashed{\partial}-\tilde e\sqrt{\hbar}\tilde{\slashed{A}}-\tilde m)\tpsi - \quart {\tilde F}_{\mu\nu}{\tilde F}^{\mu\nu}\Big]\label{qedactt}
\eeqa
where in the latter expression
\beq
\tilde\psi \equiv \psi/\sqrt{\hbar},\ \ \ \tilde A \equiv A/\sqrt{\hbar}.
\eeq
The $\hbar$ dependence now appears exclusively in the interaction term, in the combination $\tilde e\sqrt{\hbar}$.

The dimensions of the fields and parameters in the action \eq{qedact} can be easily worked out given that ${\cal S}_{QED}/\hbar$ is dimensionless, and the dimension of $\hbar$ in terms of energy $E$ and length $L$ is $[\hbar]= E\cdot L$ (for $c=1$). Thus $[\psi]=E^{1/2}\cdot L^{-1},\ [A]= E^{1/2}\cdot L^{-1/2},\ [\tilde m] = 1/L$ and $[\tilde e] = E^{-1/2}\cdot L^{-1/2}$.

The coupling $\tilde e$ appearing in the action  $\cal S$ in \eq{qedact} has a different dimension than the classical charge, $[e] = E^{1/2}\cdot L^{1/2} = [\tilde e \hbar]$. The dimensionless fine structure constant is then
\beq\label{qedalpha}
\alpha \equiv \frac{e^2}{4\pi\hbar} =\frac{{\tilde e}^2\hbar}{4\pi}
\eeq 
Each loop correction brings a factor ${\tilde e}^2$ in the amplitude and thus also one power of $\hbar$, if $\tilde e$ is independent of $\hbar$. Born (tree-level) approximations are thus obtained in the $\hbar\to 0$ limit at fixed $\tilde e$ \footnote{In a study of how classical physics emerges from quantum field theory it may be more natural to keep the classical charge $e$ fixed as $\hbar \to 0$ \cite{Holstein:2004dn}. It then turns out that classical fields get contributions also from quantum loops.}.

%
\begin{figure}[h]\label{fig4}
\includegraphics[width=12.5cm]{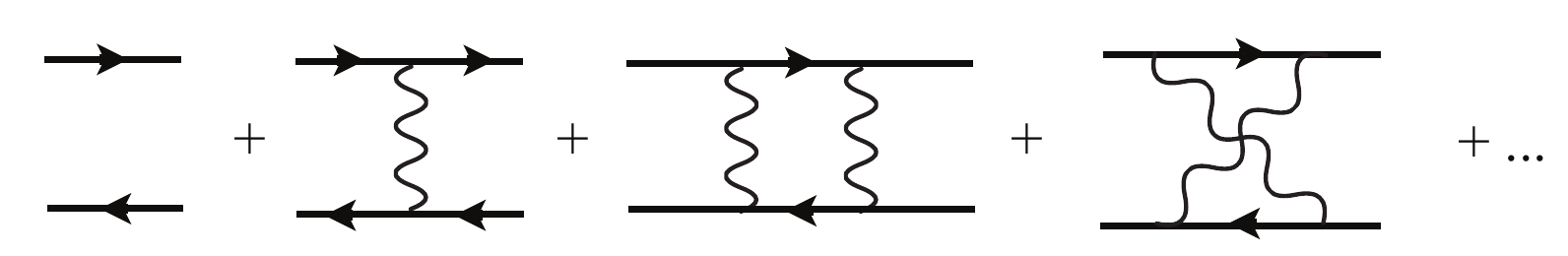}
\caption{Sum of ladder diagrams which reduces to the tree diagram sum of Fig.~3 when the antifermion mass tends to infinity.}
\end{figure}
%

The tree-level diagrams shown in Fig.~3, whose sum gives the Coulomb-Dirac bound states, stem in field theory from the crossed and uncrossed ladder diagrams shown in Fig.~4 \cite{Brodsky:1971sk}. One of the charged particles turns into the source of the Coulomb potential as its mass tends to infinity (in its rest frame). In contrast to what I said above, ladder diagrams with loops in this case give tree-level (Born) dynamics. This is possible since the bound state momenta depend on $\hbar$. Atomic binding energies are of \order{\alpha^2 m} and hence $\propto \hbar^2$ according to \eq{qedalpha}. In order to stay at the position of a bound state pole as $\hbar \to 0$ the momenta must be reduced accordingly. Propagators with loop momenta of \order{\alpha m} give inverse powers of momenta which cancel factors of $\alpha = {\tilde e}^2\hbar/4\pi$ in the numerator. Hence all ladder diagrams contribute at the same order of $\alpha$ and $\hbar$, allowing the perturbative sum to diverge at the bound state energies for any $\alpha$. Loop corrections to propagators and vertices on the other hand give genuine higher order corrections in $\alpha$ and $\hbar$.

\section{Hamiltonian formulation of Dirac states}

The QED Hamiltonian creates electron-positron pairs from the perturbative vacuum $\ket{0}$ in the presence of an external Coulomb potential. This reflects the $Z$-diagram contributions shown in Fig.~2, and implies that electron bound states must have Fock components with an unlimited number of pairs. I shall now discuss how a Hamiltonian approach without pair production may be formulated, which corresponds to the retarded propagation discussed above. In the next Section I apply it to bound states formed by mutual interactions of relativistic particles \cite{Hoyer:2009ep}. Due to the absence of loop contributions this Hamiltonian formulation is equivalent to the standard one only at the Born level.

Particle production is suppressed and retarded propagators obtained when the boundary condition is specified by the ``retarded vacuum'',
\beq\label{retvac}
\ket{0}_{R} = N^{-1}\prod_{\pv,\lambda} d_{\pv,\lambda}^\dag \ket{0}
\eeq
where the product is over all momenta $\pv$ and helicities $\lambda$ of the antifermion creation operator. The normalization factor $N$ is fixed by ${_{R}\bra{0}}0\rangle_{R} = 1$.
In the retarded vacuum
\beq\label{annret}
b_{\pv,\lambda}\ket{0}_{R}=d_{\pv,\lambda}^\dag\ket{0}_{R}=0\hspace{.5cm} {\rm and\ hence\ \ } \psi(x)\ket{0}_{R}=0
\eeq
where $\psi(x)$ is the free (interaction picture) fermion field. Consequently the retarded propagator \eq{srt} is given by the standard operator matrix element in the retarded vacuum,
\beq\label{sr2}
S_{R}(x-y)= {_{R}\bra{0}}\,T[\psi(x)\bar\psi(y)]\,\ket{0}_{R}
\eeq
The negative energy contribution to the propagator arises from the $d^\dag d$ term, which represents the removal of a positive energy antifermion from $\ket{0}_{R}$. The interaction Hamiltonian annihilates the retarded vacuum,
\beq\label{intham0}
H_{I}(t)\ket{0}_{R} = e\int d^3\xv\,A^0(\xv)\, \psi^\dag(t,\xv)\psi(t,\xv)\ket{0}_{R} =0
\eeq
which ensures the absence of particle production.

A fermion bound state at $t=0$ may be parametrized in terms of its 4-component Dirac  (c-number) wave function $\varphi(\xv)$ as
\beqa\label{dstate}
\ket{E,t=0} \equiv \int d^3\xv \psi^\dag(t=0,\xv)\varphi(\xv)\rket \hspace{5.5cm}\nn\\
= \int\dpve \sum_{\lambda}\left[u^\dag(\pv,\lambda)\varphi(\pv)b_{\pv,\lambda}^\dag\ket{0}_{R}+ v^\dag(-\pv,\lambda)\varphi(\pv)d_{-\pv,\lambda}\ket{0}_{R}\right]
\hspace{.5cm}
\eeqa
The negative energy components of $\varphi(\pv)$ describe a state where $d_{-\pv,\lambda}$ has removed a positive energy antifermion from $\ket{0}_{R}$. For the bound state to be stationary in time each Fock state amplitude $\phi(t,\xv)$ must satisfy
\beq\label{dstat}
\phi(t,\xv) \equiv \rbra\psi(t,\xv)\ket{E,t} = e^{-iEt}\phi(0,\xv)
\eeq
where $\phi(0,\xv) = \varphi(\xv)$ follows from $\acom{\psi_{\alpha}(t,\xv)}{\psi_{\beta}^\dag(t,\xv')}=\delta^3(\xv-\xv')\,\delta_{\alpha\beta}$ and $\psi(x)\rket=0$.

The time dependence of $\ket{E,t}$ is given by the interaction Hamiltonian \eq{intham0}.
The stationarity requirement for the bound state at $t=0$ is then
\beq\label{dbound}
i\frac{d\phi(0,\xv)}{dt} = \rbra i\frac{d\psi(0,\xv)}{dt}\ket{E,0} + \rbra\psi(0,\xv)H_{I}\ket{E,0} = E \phi(0,\xv)
\eeq
The interaction picture fields satisfy
\beq\label{psieom}
i\frac{d\psi(t,\xv)}{dt} = \gamma^0(-i\nv\cdot\gv +m_{e})\psi(t,\xv)
\eeq
and, making use of $\rbra \psi^\dag(t,\xv)=0$,
\beq
\rbra \psi(0,\xv)H_{I}\ket{E,0} = eA^0(\xv)\varphi(\xv)
\eeq
Using these relations in \eq{dbound} gives the Dirac equation for the wave function $\varphi(\xv)$ of a bound state of energy $E$ in the external potential $A^0(\xv)$,
\beq\label{diraceq}
(-i\nv\cdot\gv+e\gamma^0 A^0(\xv) +m_{e})\varphi(\xv) = E\gamma^0 \varphi(\xv)
\eeq

\section{Hamiltonian formulation of mesons in QCD}

\subsection{The importance of Coulomb exchange}

The above derivation of QED bound states in an external Coulomb potential, based on the retarded vacuum $\rket$, gave the correct (Dirac) bound state energy and quantum numbers at the Born level. I now discuss how this method may be applied to Born level meson ($\qu\qb$) bound states in QCD. The wave function will describe a two-body state, whereas, like in the Dirac case, the same bound state formulated on the standard vacuum $\ket{0}$ has an indefinite number of sea quark pairs.

The instantaneous Coulomb potential is determined by the positions of the charged constituents through Gauss' law. Since transverse gluons propagate in time their exchange involves intermediate states which include gluons. The derivation below concerns the valence $\qu\qb$ Fock states only and thus applies insofar as Coulomb exchange is dominant. This is trivially the case in $D=1+1$ dimensions, where there are no transverse degrees of freedom. However, the derivation applies also in $D=3+1$ dimensions provided one allows for a homogeneous solution of Gauss' law (Section \ref{linsec} below). This gives rise to a linear Coulomb potential which is of lower order in $\alpha_s$ compared to the contribution of transverse gluon exchange. The linear potential is of course doubly welcome as a way of providing confinement.

\subsection{The $u\bar d$ wave function $\chi$ and its gauge dependence}

The $u\bar d$ bound state at $t=0$ is analogously to \eq{dstate} expressed as
\beq\label{ffqcd}
\ket{E,t=0}=\int d^3\yv_{1}d^3\yv_{2}\,\psi_{u}^{A\dag}(t=0,\yv_{1})\chi^{AB}(\yv_{1},\yv_{2})\psi_{d}^{B}(t=0,\yv_{2})\ket{0}_{R}
\eeq
where now
\beq\label{retvac3}
\ket{0}_{R} = N^{-1}\prod_{\pv,\lambda,A} d_{u}^{A\dag}(\pv,\lambda)\, b_{d}^{A\dag}(\pv,\lambda) \ket{0}
\eeq
includes a product over colors.
The wave function $\chi$ which describes the state \eq{ffqcd} is gauge dependent. Invariance of $\ket{E,t}$ under time-independent gauge transformations $\psi(t,\xv) \to U(\xv)\psi(t,\xv)$ implies
\beq
\chi(\yv_{1},\yv_{2}) \to U(\yv_{1})\chi(\yv_{1},\yv_{2}) U^{\dag}(\yv_{2})
\eeq
As an {\it ansatz} for a bound state solution I assume the existence of a gauge where the wave function is diagonal in color,
\beq\label{chising}
\chi^{AB}(\yv_{1},\yv_{2}) = \delta^{AB} \chi(\yv_{1},\yv_{2})
\eeq
The stationarity condition for the bound state wave function corresponding to \eq{dstat} is
\beq\label{bsqcd}
\phi_{\alpha\beta}^{CD}(t;\xv_{1},\xv_{2}) \equiv \rbra\psi_{d\beta}^{D\dag}(t,\xv_{2}) \psi_{u\alpha}^{C}(t,\xv_{1})\ket{E,t} = e^{-iEt} \phi_{\alpha\beta}^{CD}(t=0;\xv_{1},\xv_{2})
\eeq
where $\phi_{\alpha\beta}(t=0;\xv_{1},\xv_{2})= \chi_{\alpha\beta}(\xv_{1},\xv_{2})$. 

\subsection{The $\qu\qb$ matrix element of the gluon equations of motion}

The gluon field is determined by the equations of motion {\it for each Fock state,} \ie, for each $\qu\qb$ configuration with the $u$-quark of color $C$ located at $\xv_1$ and the $\bar d$-antiquark of color $D$ at $\xv_2$. The corresponding matrix element of the equation of motion,
\beqa\label{bseom}
\rbra\psi_{d\beta}^{D\dag}(t,\xv_{2}) \psi_{u\alpha}^{C}(t,\xv_{1})\hspace{7cm}\nn\\
\times \Big[\partial_{\mu}F_{a}^{\mu\nu}+gf_{abc}F_{b}^{\mu\nu}A_{\mu}^c - g\sum_{f=u,d}\bar\psi_{f}^{A}\gamma^\nu T_{a}^{AB}\psi_{f}^{B}\Big]\ket{E,t} =0
\eeqa
should vanish to leading order in $g$. This gives
\beqa\label{bseom2}
\chi^{CD}(\xv_{1},\xv_{2}) \left[\partial_{\mu}F_{a}^{\mu\nu}+gf_{abc}F_{b}^{\mu\nu}A_{\mu}^c\right] &=& g\,\delta^3(\xv-\xv_{1})T_{a}^{CA}\gamma^0\gamma^\nu \chi^{AD}(\xv_{1},\xv_{2})\nn\\[3mm]
&-& g\,\delta^3(\xv-\xv_{2})\chi^{CA}(\xv_{1},\xv_{2})\gamma^0\gamma^\nu T_{a}^{AD} \nn\\[2mm]
\eeqa
Only the color diagonal $a=3,8$ components of the $A_{a}^\mu$ fields can be non-vanishing in the gauge where the wave function has the color structure \eq{chising}. Then the commutator contributions vanish and the solution is abelian.

In $D=1+1$ the Coulomb potential $A_{3,8}^0$ is linear and we may set $A_{3,8}^1=0$. The bound state wave functions can then be derived as in Section \ref{sectbse} below. 

\subsection{The linear potential in 3+1 dimensions}\label{linsec}

The field equations are consistent with an instantaneous linear potential even in $D=3+1$ dimensions. The homogeneous solution
\beq\label{homsol}
A^0_{a}(\xv) = \la^2_{a}\,\chat_{a}\cdot\xv \hspace{1cm} (a=3,8)
\eeq 
trivially satisfies $\nv^2 A^0_{a} = 0$ for any ($\xv$-independent) values of the parameters $\la_a$ and unit vectors $\chat_a$. Being of \order{g^0} this solution dominates the \order{g} fields generated by the quark sources. Hence we may set $\bs{A}_a=0$ at leading order in $g$, ignoring transverse gluon exchange and the higher Fock components this would entail. Note that this is true also for relativistic constituents and in all coordinate frames.

The fact that \eq{homsol} (with $\bs{A}_a=0$) solves the field equations \eq{bseom2} at \order{g^0} implies that the action is stationary under {\it local} variations of the fields. The action should also be stationary under variations of the {\it global} parameters $\la_{a},\,\chat_{a}$ of the potential \eq{homsol}. This will ensure that the solution has rotational and color symmetry.

The lagrangian density $F_{\mu\nu}^{a}F^{\mu\nu}_{a}$ has an \order{g^0} contribution from the square of the homogeneous solution \eq{homsol}, an \order{g} term from the interference of the homogeneous solution with the $A^0_{a=3,8}$ fields generated by the quark sources in the field equations \eq{bseom2}, and \order{g^2} contributions from squares of the \order{g} fields. We need to consider the action at \order{g} and thus also the field equations for $A^0_{a=3,8}$ including the \order{g} quark sources,
\beq\label{bseom3}
-\nv^2 A_{a}^{0}(\xv) = g\,T_{a}^{CC}\left[\delta^3(\xv-\xv_{1})
- \delta^3(\xv-\xv_{2})\right] \hspace{1cm} (a=3,8)
\eeq
(no sum on the quark color $C$). The solution
\beq\label{a0qcd}
A^0_{a}(\xv;\xv_{1},\xv_{2},C) = \la^2_{a}\,\chat_{a}\cdot\xv+\frac{gT^{CC}_{a}}{4\pi}\left(\inv{|\xv-\xv_{1}|}-\inv{|\xv-\xv_{2}|}\right) \hspace{1cm} (a = 3,8)
\eeq
depends on the positions and color of the quarks, as well as on the parameters $\la_{a},\, \chat_{a}$ of the homogeneous solution. The instantaneous gluon action with this Coulomb field is
\beqa\label{lingauge2}
-\inv{4}\sum_{a}\int d^3\xv F_{\mu\nu}^{a}F^{\mu\nu}_{a} = \inv{2}\sum_{a}\int d^3\xv\, (\nv A_{a}^0)^2 \hspace{4cm}\\
=\sum_{a=3,8}\left[\inv{2}\la_{a}^4\int d^3\xv + \inv{3} g\Lambda_{a}^2\, T_{a}^{CC}\chat_{a}\cdot (\xv_{1}-\xv_{2}) +\morder{g^2}\right] \nn
\eeqa
The parameter $\sum_{a=3,8}\la_{a}^4$ is multiplied by the (infinite) volume of space. This term does not affect bound state evolution provided it is the {\it same} for all $\qu\qb$ configurations. Hence we must require that
\beq\label{lamdef}
\la^4 \equiv\sum_{a=3,8}\la_{a}^4 
\eeq
is a universal constant, independent of $\xv_{1},\xv_{2}$ and the quark color $C$. The \order{g} interference term in \eq{lingauge2} is stationary \wrt variations of the unit vectors $\chat_{a}$ provided $\chat_{a}\parallel \xv_{1}-\xv_{2}$. Choosing $\chat_{a}=T_{a}^{CC}(\xv_{1}-\xv_{2})/|T_{a}^{CC}(\xv_{1}-\xv_{2})|$ gives (as seen below) an {\it attractive} linear potential. 

The (instantaneous) action \eq{lingauge2} should be stationary also \wrt variations in the ratio $\la_{3}/\la_{8}$ which leaves $\la$ in \eq{lamdef} invariant. For quarks of color $C=1$ the extremum of the \order{g} term in \eq{lingauge2} is obtained for $\la_{3}^2/\la_{8}^2=\sqrt{3}$:
\beq\label{sint1}
S_{int}^{C=1} = \max\left\{\frac{g}{6}\left(\la_{3}^2+\inv{\sqrt{3}}\la_{8}^2\right)|\xv_{1}-\xv_{2}|\right\} = \frac{g\la^2}{3\sqrt{3}}|\xv_{1}-\xv_{2}|
\eeq
The calculation is the same for $C=2$, whereas for $C=3$ the extremum is obtained for $\la_{3}^2/\la_{8}^2=0$:
\beq\label{sint3}
S_{int}^{C=3} = \max\left\{\frac{g\la_{8}^2}{3\sqrt{3}}|\xv_{1}-\xv_{2}|\right\} = \frac{g\la^2}{3\sqrt{3}}|\xv_{1}-\xv_{2}|
\eeq
The factorization of the color dependence as in the {\it ansatz} \eq{chising} requires that the interactions are the same for all color components, \ie, that $S_{int}^{C=3}=S_{int}^{C=1}\equiv S_{int}$. The equality of \eq{sint1} and \eq{sint3} was ensured by the color singlet nature of the action \eq{lingauge2} and the color covariance of the equations of motion.

\subsection{Bound state equation}\label{sectbse}

Having determined the parameters $\la_{a}$ and $\chat_{a}$ in the $A_{a}^0$ potential \eq{a0qcd} for each Fock state I proceed to impose the stationary time dependence \eq{bsqcd} on the bound state. Analogously to the Dirac case \eq{dbound} I get 
\beqa\label{bsamp}
i\frac{d\phi_{\alpha\beta}^{CD}(0;\xv_{1},\xv_{2})}{dt}= E\,\phi_{\alpha\beta}^{CD}(0;\xv_{1},\xv_{2}) =\hspace{5cm}\\ 
=\rbra i\frac{d\psi_{d\beta}^{D\dag}(0,\xv_{2})}{dt} \psi_{u\alpha}^{C}(0,\xv_{1})\ket{E,0} + i\psi_{d\beta}^{D\dag}(0,\xv_{2})\frac{d\psi_{u\alpha}^{C}(0,\xv_{1})}{dt}\ket{E,0}\nn\\
 +\rbra\psi_{d\beta}^{D\dag}(0,\xv_{2}) \psi_{u\alpha}^{C}(0,\xv_{1})[H_{I}(0)-S_{int}]\ket{E,0}\nn
\eeqa
Here the Coulomb energy $-S_{int}$ stored in the gluon field contributes to the Hamiltonian with the opposite sign compared to the action \eq{lingauge2}. Using the time dependence \eq{psieom} of the fields, the color structure \eq{chising} of the wave function and the interaction hamiltonian
\beq\label{hiqcd}
H_{I}(t) = g\sum_{f=u,d}\int d^3\xv \psi_{f}^{A\dag}(t,\xv)A_{a}^0(\xv)\, T_{a}^{AB}\psi_{f}^{B}(t,\xv)
\eeq
gives the bound state equation
\beqa\label{ffbse}
\gamma^0(-i\nv_{1}\cdot\gv +m_{u})\chi(\xv_{1},\xv_{2}) - \chi(\xv_{1},\xv_{2})\gamma^0(i\nv_{2}\cdot\gv +m_{d})&&\nn\\[3mm] = [E-V(\xv_{1},\xv_{2})] \chi(\xv_{1},\xv_{2})&&
\eeqa
Not surprisingly, it has the form of a ``double Dirac equation'' and as such was proposed a long time ago by Breit \cite{Breit:1929zz}. However, the approximations required to derive this equation from the underlying gauge theory have been clarified: lowest order in $\hbar$ (allowing the use of retarded propagation) as well as lowest order in the coupling $g$. The potential must be linear,
\beq\label{mespot}
V(\xv_{1},\xv_{2}) = \frac{2g\la^2}{3\sqrt{3}}|\xv_{1}-\xv_{2}|
\eeq
where $\la$ is a free parameter with dimension of mass. In $D=1+1$ dimensions even the perturbative potential is linear, with a coefficient $\propto g^2$. In $D=3+1$ the linear potential is obtained through a boundary condition for Gauss' law characterized by $\la$ through Eqs. \eq{homsol} and \eq{lamdef}.

\section{Discussion}

I have addressed an apparent paradox of hadron structure sketched in Fig.~1: The quantum numbers of hadrons reflect only their valence degrees of freedom ($\qu\qb$ or $\qu\qu\qu$) even though direct DIS measurements indicate that there is a sea of quark pairs. This feature is closely related to the relativistic nature of hadrons, and is seen also for an electron in an external Coulomb potential. Dirac bound states have an infinite number of Fock components due to pair production in the perturbative vacuum. The same bound states are obtained at Born (no loop) level in a retarded vacuum, which suppresses pair production. The wave function then describes a single ``valence'' electron with both positive and negative energy components: The standard Dirac wave function. This example is encouraging as it shows that quantum numbers and mass spectra of relativistic bound states with an infinite number of constituents can be addressed analytically.

Tree-level interactions are insensitive to the $\ieps$ prescription of propagators, allowing the use of retarded boundary conditions. The basic approximation may thus be formulated as being of lowest order in $\hbar$, \ie, a Born approximation for bound states \cite{Hoyer:2009ep,Brodsky:2010zk}. This approximation is valid for relativistic dynamics and provides a well defined starting point in the study of hadron structure.

Another issue arises when the quark interactions are determined from the field equations rather than imposed by an external potential. Relativistically moving charges generate transverse gauge fields which propagate with finite speed. Hence transverse gluon exchange involves intermediate Fock states with gluon constituents. This was demonstrated in QED for the Hydrogen atom \cite{Jarvinen:2004pi}. Part of the binding energy of atoms in relativistic CM motion arises from transverse photon exchange and thus involves $\ket{ep\gamma}$ Fock states in addition to the $\ket{ep}$ state which dominates in the rest frame.

A valence ($\qu\qb$ or $\qu\qu\qu$) description of hadrons is thus only possible when the instantaneous Coulomb potential dominates transverse exchange. In $D=1+1$ dimensions this is trivially true since there are no transverse gauge fields. In $3+1$ dimensions the Coulomb field also dominates provided one allows for a homogeneous, \order{g^0} solution of Gauss' law. Such a solution gives a constant color field extending to infinity and would seem to be unphysical. However, the instantaneous field is determined separately for each quark configuration and thus {\it depends on the position and color of the constituents}. The combined field strength far from color singlet hadrons in fact vanishes due to the coherent sum over quark colors. Translation invariance moreover requires that the hadrons be color singlets.

The linear potential is of course very welcome also since it provides color confinement. I should emphasize that my analysis only shows that a linear solution is {\it consistent} with QCD at leading order. It does not provide a theoretical argument for why such a boundary condition must be used in QCD. To this day we also lack a proof that QED does {\it not} confine -- which has not impeded theoretical progress.

An analysis similar to the one outlined above for mesons can be carried out also for baryons \cite{Hoyer:2009ep}. The potential turns out to be
\beq\label{barpot}
V(\xv_{1},\xv_{2},\xv_{3})=\frac{\sqrt{2}g\la^2}{3\sqrt{3}}\,\sqrt{(\xv_{1}-\xv_{2})^2+(\xv_{2}-\xv_{3})^2+(\xv_{3}-\xv_{1})^2}
\eeq
when the three quarks are located at $\xv_{1},\xv_{2},\xv_{3}$ and the color dependence of the wave function is given by $\varepsilon^{ABC}$. In the limit where two of the quarks are close together the potential \eq{barpot} approaches the linear one \eq{mespot} \wrt the position of the third quark. This construction is only possible when the number of colors equals the number of valence quarks.

If the CM momentum $\kv$ in the $u\bar d$ meson state \eq{ffqcd} is separated as
\beq\label{cmsep}
\chi(\xv_{1},\xv_{2}) = e^{i\kv\cdot(\xv_{1}+\xv_{2})/2}\,\chi_{\kv}(\xv_{1}-\xv_{2})
\eeq
the bound state equation reduces to
\beqa\label{bse}
-i\nv\cdot\com{\bs{\alpha}}{\chi_{\kv}(\xv)}+\halft\kv\cdot\acom{\bs{\alpha}}{\chi_{\kv}(\xv)}+m_{u}\gamma^0\chi_{\kv}(\xv)-\chi_{\kv}(\xv)\gamma^0 m_{d}\nn\\[3mm] = (E_{\kv}-V)\chi_{\kv}(\xv)
\eeqa
Because the bound state constituents are at equal time in all frames the $\kv$-dependence of the wave function $\chi_{\kv}$ is dynamical. However, the $\kv$-dependence of the bound state energy must be kinematic, $E_{\kv}=\sqrt{\kv^2+M^2}$. Remarkably, the bound state equation \eq{bse} gives precisely this relation -- and only for a linear potential \cite{Hoyer:1985tz}. The wave function Lorentz contracts in an intriguing way, since the boost parameter depends on the distance $|\xv|$ between the quarks, $\tanh\zeta=|\kv|/(E_{\kv}-V(|\xv|))$. Non-relativistic bound states ($E_{\kv}\gg V$) Lorentz contract in the standard way. 

The rest frame solutions of Eq. \eq{bse} have been studied using phenomenological potentials \cite{Geffen:1977bh,Krolikowski:1992fy}. The radial wave functions can be singular at $r=0$ and at $E-V(r)=0$. Solutions that are regular at both points have quantized masses which lie on asymptotically linear Regge trajectories. The wave function oscillates rapidly at large distances $r$ between the quarks, where $V(r) \gg E$. In this region the normalization per unit distance is constant, which may reflect the virtual pairs of the color string.

The linear confining potential \eq{mespot} appears as a ``zeroth'' order approximation in a perturbative expansion. Perturbative contributions of \order{\as} from gluon exchange, quark annihilation and loops still need to be added. This is consistent with the spirit of quark models, which typically assume a potential of the form
\beq\label{qmodel}
V(r)=C\, r-C_F\frac{\alpha_s}{r}
\eeq
where a non-perturbative linear potential is added to perturbative gluon exchange. This can be self-consistent provided $\as$ is not too large even at the confinement scale $Q^2=C$, and if the linear potential may be effectively treated as a lowest order term in a perturbative expansion.

\vspace{1cm} \noindent
{\bf Acknowledgements}

I am grateful and honored for the invitation to participate in this 50th Cracow School of Theoretical Physics. A long-time collaboration with Stan Brodsky on issues related to bound states is gratefully acknowledged, as well as helpful remarks by St\'ephane Peign\'e.

\end{document}